\documentclass[twocolumn,aps,prl,10pt,floatfix]{revtex4-2}
\usepackage[utf8]{inputenc}
\usepackage{amsmath}
\usepackage{amsfonts}
\usepackage{amssymb}
\usepackage{graphicx}
\usepackage{hyperref}
\usepackage{braket}

\begin{document}

\title{Quantum phase transition from a superfluid to a Mott insulator\\ in a gas of ultracold atoms}

\author{
  Markus Greiner$^{1,2}$, 
  Olaf Mandel$^{1,2}$, 
  Tilman Esslinger$^{3}$, 
  Theodor W. Hänsch$^{1,2}$, 
  Immanuel Bloch$^{1,2}$\thanks{\email{imb@mpq.mpg.de}}
}
\affiliation{$^{1}$Sektion Physik, Ludwig-Maximilians-Universität, Schellingstrasse 4/III, D-80799 Munich, Germany}
\affiliation{$^{2}$Max-Planck-Institut für Quantenoptik, D-85748 Garching, Germany}
\affiliation{$^{3}$Quantenelektronik, ETH Zürich, 8093 Zurich, Switzerland}

\begin{abstract}
For a system at a temperature of absolute zero, all thermal fluctuations are frozen out, while quantum fluctuations prevail. These microscopic quantum fluctuations can induce a macroscopic phase transition in the ground state of a many-body system when the relative strength of two competing energy terms is varied across a critical value. Here we observe such a quantum phase transition in a Bose–Einstein condensate with repulsive interactions, held in a three-dimensional optical lattice potential. As the potential depth of the lattice is increased, a transition is observed from a superfluid to a Mott insulator phase. In the superfluid phase, each atom is spread out over the entire lattice, with long-range phase coherence. But in the insulating phase, exact numbers of atoms are localized at individual lattice sites, with no phase coherence across the lattice; this phase is characterized by a gap in the excitation spectrum. We can induce reversible changes between the two ground states of the system.
\end{abstract}

\maketitle

\section{Introduction}

A physical system that crosses the boundary between two phases changes its properties in a fundamental way. It may, for example, melt or freeze. This macroscopic change is driven by microscopic fluctuations. When the temperature of the system approaches zero, all thermal fluctuations die out. This prohibits phase transitions in classical systems at zero temperature, as their opportunity to change has vanished. However, their quantum mechanical counterparts can show fundamentally different behaviour. In a quantum system, fluctuations are present even at zero temperature, due to Heisenberg's uncertainty relation. These quantum fluctuations may be strong enough to drive a transition from one phase to another, bringing about a macroscopic change.

A prominent example of such a quantum phase transition is the change from the superfluid phase to the Mott insulator phase in a system consisting of bosonic particles with repulsive interactions hopping through a lattice potential. This system was first studied theoretically in the context of superfluid-to-insulator transitions in liquid helium~\cite{fisher1989}. Recently, Jaksch et al.~\cite{jaksch1998} have proposed that such a transition might be observable when an ultracold gas of atoms with repulsive interactions is trapped in a periodic potential.

To illustrate this idea, we consider an atomic gas of bosons at low enough temperatures that a Bose–Einstein condensate is formed. The condensate is a superfluid, and is described by a wavefunction that exhibits long-range phase coherence~\cite{stringari2001}. An intriguing situation appears when the condensate is subjected to a lattice potential in which the bosons can move from one lattice site to the next only by tunnel coupling. If the lattice potential is turned on smoothly, the system remains in the superfluid phase as long as the atom–atom interactions are small compared to the tunnel coupling. In this regime a delocalized wavefunction minimizes the dominant kinetic energy, and therefore also minimizes the total energy of the many-body system. In the opposite limit, when the repulsive atom–atom interactions are large compared to the tunnel coupling, the total energy is minimized when each lattice site is filled with the same number of atoms. The reduction of fluctuations in the atom number on each site leads to increased fluctuations in the phase. Thus in the state with a fixed atom number per site phase coherence is lost. In addition, a gap in the excitation spectrum appears. The competition between two terms in the underlying hamiltonian (here between kinetic and interaction energy) is fundamental to quantum phase transitions~\cite{sachdev2001} and inherently different from normal phase transitions, which are usually driven by the competition between inner energy and entropy.

The physics of the above-described system is captured by the Bose–Hubbard model~\cite{fisher1989}, which describes an interacting boson gas in a lattice potential. The hamiltonian in second quantized form reads:

\begin{equation}
H = -J \sum_{\langle i,j \rangle} \hat{a}^\dagger_i \hat{a}_j + \sum_i \epsilon_i \hat{n}_i + \frac{1}{2} U \sum_i \hat{n}_i(\hat{n}_i - 1)
\end{equation}

Here $\hat{a}^\dagger_i$ and $\hat{a}_i$ correspond to the bosonic annihilation and creation operators of atoms on the $i$th lattice site, $\hat{n}_i = \hat{a}^\dagger_i \hat{a}_i$ is the atomic number operator counting the number of atoms on the $i$th lattice site, and $\epsilon_i$ denotes the energy offset of the $i$th lattice site due to an external harmonic confinement of the atoms~\cite{jaksch1998}.

The strength of the tunnelling term in the hamiltonian is characterized by the hopping matrix element between adjacent sites $i,j$:
\begin{equation}
J = -\int d^3x \, w^*(\mathbf{x}-\mathbf{x}_i) \left[-\frac{\hbar^2}{2m} \nabla^2 + V_{lat}(\mathbf{x})\right] w(\mathbf{x}-\mathbf{x}_j),
\end{equation}
where $w(\mathbf{x}-\mathbf{x}_i)$ is a single particle Wannier function localized to the $i$th lattice site (as long as $n_i \approx O(1))$, $V_{lat}(\mathbf{x})$ indicates the optical lattice potential and $m$ is the mass of a single atom.

The repulsion between two atoms on a single lattice site is quantified by the on-site interaction matrix element:
\begin{equation}
U = \frac{4\pi\hbar^2 a}{m} \int |w(\mathbf{x})|^4 d^3x
\end{equation}
with $a$ being the scattering length of an atom. In our case the interaction energy is very well described by the single parameter $U$, due to the short range of the interactions, which is much smaller than the lattice spacing.

In the limit where the tunnelling term dominates the hamiltonian, the ground-state energy is minimized if the single-particle wavefunctions of $N$ atoms are spread out over the entire lattice with $M$ lattice sites. The many-body ground state for a homogeneous system ($\epsilon_i = \text{const.}$) is then given by:

\begin{equation}
\ket{\Psi_{SF}}_{U=0} \propto \left(\sum_{i=1}^M \hat{a}^\dagger_i\right)^N \ket{0}
\end{equation}

Here all atoms occupy the identical extended Bloch state. An important feature of this state is that the probability distribution for the local occupation $n_i$ of atoms on a single lattice site is poissonian, that is, its variance is given by $\text{Var}(n_i) = \langle \hat{n}_i \rangle$. Furthermore, this state is well described by a macroscopic wavefunction with long-range phase coherence throughout the lattice.

If interactions dominate the hamiltonian, the fluctuations in atom number of a Poisson distribution become energetically very costly and the ground state of the system will instead consist of localized atomic wavefunctions with a fixed number of atoms per site that minimize the interaction energy. The many-body ground state is then a product of local Fock states for each lattice site. In this limit, the ground state of the many-body system for a commensurate filling of $n$ atoms per lattice site in the homogeneous case is given by:

\begin{equation}
\ket{\Psi_{MI}}_{J=0} \propto \prod_{i=1}^M \left(\hat{a}^\dagger_i\right)^n \ket{0}
\end{equation}

This Mott insulator state cannot be described by a macroscopic wavefunction like in a Bose condensed phase, and thus is not amenable to a treatment via the Gross-Pitaevskii equation or Bogoliubov's theory of weakly interacting bosons. In this state no phase coherence is prevalent in the system, but perfect correlations in the atom number exist between lattice sites.

As the strength of the interaction term relative to the tunnelling term in the Bose–Hubbard hamiltonian is changed, the system reaches a quantum critical point in the ratio of $U/J$, for which the system will undergo a quantum phase transition from the superfluid state to the Mott insulator state. In three dimensions, the phase transition for an average number of one atom per lattice site is expected to occur at $U/J \approx z \times 5.8$~\cite{fisher1989,sheshadri1993, freericks1995,oosten2001}, with $z$ being the number of next neighbours of a lattice site.

The qualitative change in the ground-state configuration below and above the quantum critical point is also accompanied by a marked change in the excitation spectrum of the system. In the superfluid regime, the excitation spectrum is gapless whereas the Mott insulator phase exhibits a gap in the excitation spectrum~\cite{sheshadri1993,freericks1995,oosten2001,elstner1999}. An essential feature of a quantum phase transition is that this energy gap $\Delta$ opens up as the quantum critical point is crossed.

Studies of the Bose–Hubbard hamiltonian have so far included granular superconductors~\cite{orr1986,haviland1989} and one- and two-dimensional Josephson junction arrays~\cite{bradley1984,geerligs1989,zwerger1989,vanderzant1992,vanoudenaarden1996,chow1998}. In the context of ultracold atoms, atom number squeezing has very recently been demonstrated with a Bose–Einstein condensate in a one-dimensional optical lattice~\cite{orzel2001}. The above experiments were mainly carried out in the limit of large boson occupancies $n_i$ per lattice site, for which the problem can be well described by a chain of Josephson junctions.

In our present experiment we load $^{87}$Rb atoms from a Bose–Einstein condensate into a three-dimensional optical lattice potential. This system is characterized by a low atom occupancy per lattice site of the order of $\langle \hat n_i \rangle \approx 1-3$, and thus provides a unique testing ground for the Bose–Hubbard model. As we increase the lattice potential depth, the hopping matrix element $J$ decreases exponentially but the on-site interaction matrix element $U$ increases. We are thereby able to bring the system across the critical ratio in $U/J$, such that the transition to the Mott insulator state is induced.

\section{Experimental technique}

The experimental set-up and procedure to create $^{87}$Rb Bose–Einstein condensates are similar to those in our previous experimental work~\cite{greiner2001a,greiner2001b}. In brief, spin-polarized samples of laser-cooled atoms in the $(F=2, m_F=2)$ state are transferred into a cigar-shaped magnetic trapping potential with trapping frequencies of $\nu_{\text{radial}} = 240$\,Hz and $\nu_{\text{axial}} = 24$\,Hz. Here $F$ denotes the total angular momentum and $m_F$ the magnetic quantum number of the state. Forced radio-frequency evaporation is used to create Bose–Einstein condensates with up to $2 \times 10^5$ atoms and no discernible thermal component. The radial trapping frequencies are then relaxed over a period of 500 ms to $\nu_{rad} = 24$\,Hz such that a spherically symmetric Bose–Einstein condensate with a Thomas–Fermi diameter of 26\,$\mu$m is present in the magnetic trapping potential.

In order to form the three-dimensional lattice potential, three optical standing waves are aligned orthogonal to each other, with their crossing point positioned at the centre of the Bose–Einstein condensate. Each standing wave laser field is created by focusing a laser beam to a waist of 125\,$\mu$m at the position of the condensate. A second lens and a mirror are then used to reflect the laser beam back onto itself, creating the standing wave interference pattern. The lattice beams are derived from an injection seeded tapered amplifier and a laser diode operating at a wavelength of $\lambda = 852$\,nm. All beams are spatially filtered and guided to the experiment using optical fibres. Acousto-optical modulators are used to control the intensity of the lattice beams and introduce a frequency difference of about 30\,MHz between different standing wave laser fields. The polarization of a standing wave laser field is chosen to be linear and orthogonal polarized to all other standing waves. Due to the different frequencies in each standing wave, any residual interference between beams propagating along orthogonal directions is time-averaged to zero and therefore not seen by the atoms. The resulting three-dimensional optical potential (see ref.\,\cite{grimm2000} and references therein) for the atoms is then proportional to the sum of the intensities of the three standing waves, which leads to a simple cubic type geometry of the lattice:
\begin{equation}
V(x,y,z) = V_0\left[\sin^2(kx) + \sin^2(ky) + \sin^2(kz)\right]
\end{equation}

Here $k = 2\pi/\lambda$ denotes the wavevector of the laser light and $V_0$ is the maximum potential depth of a single standing wave laser field. This depth $V_0$ is conveniently measured in units of the recoil energy $E_r = \hbar^2 k^2/2m$. The confining potential for an atom on a single lattice site due to the optical lattice can be approximated by a harmonic potential with trapping frequencies $\nu_r$ on the order of $\nu_r \approx \left (\hbar k^2/2\pi m \right ) \sqrt{V_0/E_r}$. In our set-up potential depths of up to 22\,$E_r$ can be reached, resulting in trapping frequencies of approximately $\nu_r \approx 30$\,kHz. The gaussian intensity profile of the laser beams at the position of the condensate creates an additional weak isotropic harmonic confinement over the lattice, with trapping frequencies of 65 Hz for a potential depth of 22\,$E_r$.

The magnetically trapped condensate is transferred into the optical lattice potential by slowly increasing the intensity of the lattice laser beams to their final value over a period of 80 ms using an exponential ramp with a time constant of $\tau = 20$ ms. The slow ramp speed ensures that the condensate always remains in the many-body ground state of the combined magnetic and optical trapping potential. After raising the lattice potential the condensate has been distributed over more than 150,000 lattice sites ($\sim$65 lattice sites in a single direction) with an average atom number of up to 2.5 atoms per lattice site in the centre.

\begin{figure}[htbp]
\centering
  \includegraphics[width=\columnwidth]{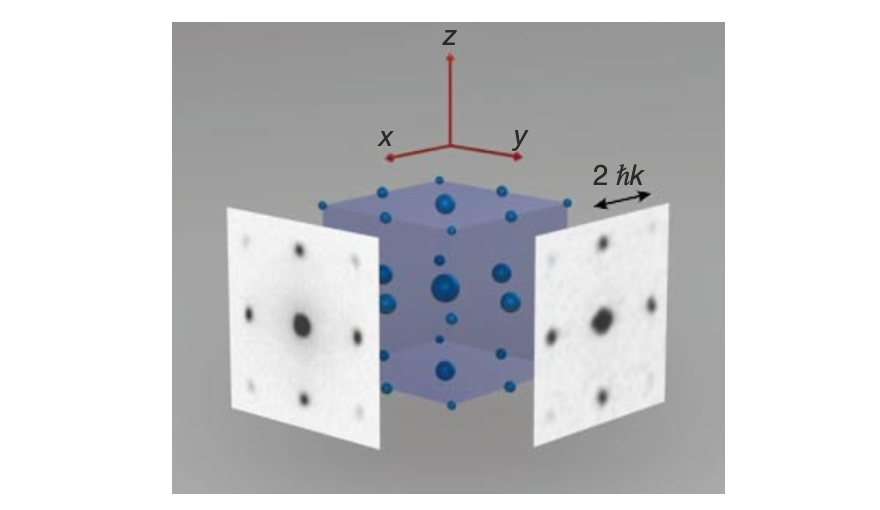} 
\caption{Schematic three-dimensional interference pattern with measured absorption images taken along two orthogonal directions. The absorption images were obtained after ballistic expansion from a lattice with a potential depth of $V_0 = 10\,E_r$ and a time of flight of 15\,ms.}
\label{fig:interference_3d}
\end{figure}

In order to test whether there is still phase coherence between different lattice sites after ramping up the lattice potential, we suddenly turn off the combined trapping potential. The atomic wavefunctions are then allowed to expand freely and interfere with each other. In the superfluid regime, where all atoms are delocalized over the entire lattice with equal relative phases between different lattice sites, we obtain a high-contrast three-dimensional interference pattern as expected for a periodic array of phase coherent matter wave sources (see Fig.~\ref{fig:interference_3d}). It is important to note that the sharp interference maxima directly reflect the high degree of phase coherence in the system for these experimental values.

\subsection{Entering the Mott insulator phase}

\begin{figure}[htbp]
\centering
  \includegraphics[width=\linewidth]{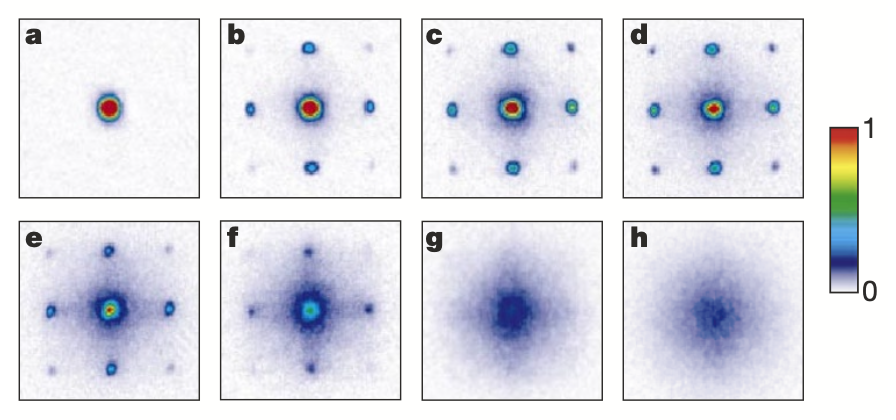} 
\caption{Absorption images of multiple matter wave interference patterns. These were obtained after suddenly releasing the atoms from an optical lattice potential with different potential depths $V_0$ after a time of flight of 15 ms. Values of $V_0$ were: \textbf{a}, $0\,E_r$; \textbf{b}, $3\,E_r$; \textbf{c}, $7\,E_r$; \textbf{d}, $10\,E_r$; \textbf{e}, $13\,E_r$; \textbf{f}, $14\,E_r$; \textbf{g}, $16\,E_r$; and \textbf{h}, $20\,E_r$.}
\label{fig:interference_evolution}
\end{figure}

As we increase the lattice potential depth, the resulting interference pattern changes markedly (see Fig.~\ref{fig:interference_evolution}). Initially the strength of higher-order interference maxima increases as we raise the potential height, due to the tighter localization of the atomic wavefunctions at a single lattice site. Quite unexpectedly, however, at a potential depth of around 13\,$E_r$ the interference maxima no longer increase in strength (see Fig.\,\ref{fig:interference_evolution}e): instead, an incoherent background of atoms gains more and more strength until at a potential depth of 22\,$E_r$ no interference pattern is visible at all. Phase coherence has obviously been completely lost at this lattice potential depth. A remarkable feature during the evolution from the coherent to the incoherent state is that when the interference pattern is still visible no broadening of the interference peaks can be detected until they completely vanish in the incoherent background. This behaviour can be explained on the basis of the superfluid–Mott insulator phase diagram. After the system has crossed the quantum critical point $U/J = z \times 5.8$, it will evolve in the inhomogeneous case into alternating regions of incoherent Mott insulator phases and coherent superfluid phases~\cite{jaksch1998}, where the superfluid fraction continuously decreases for increasing ratios $U/J$.

\begin{figure}[htbp]
\centering
  \includegraphics[width=0.8\columnwidth]{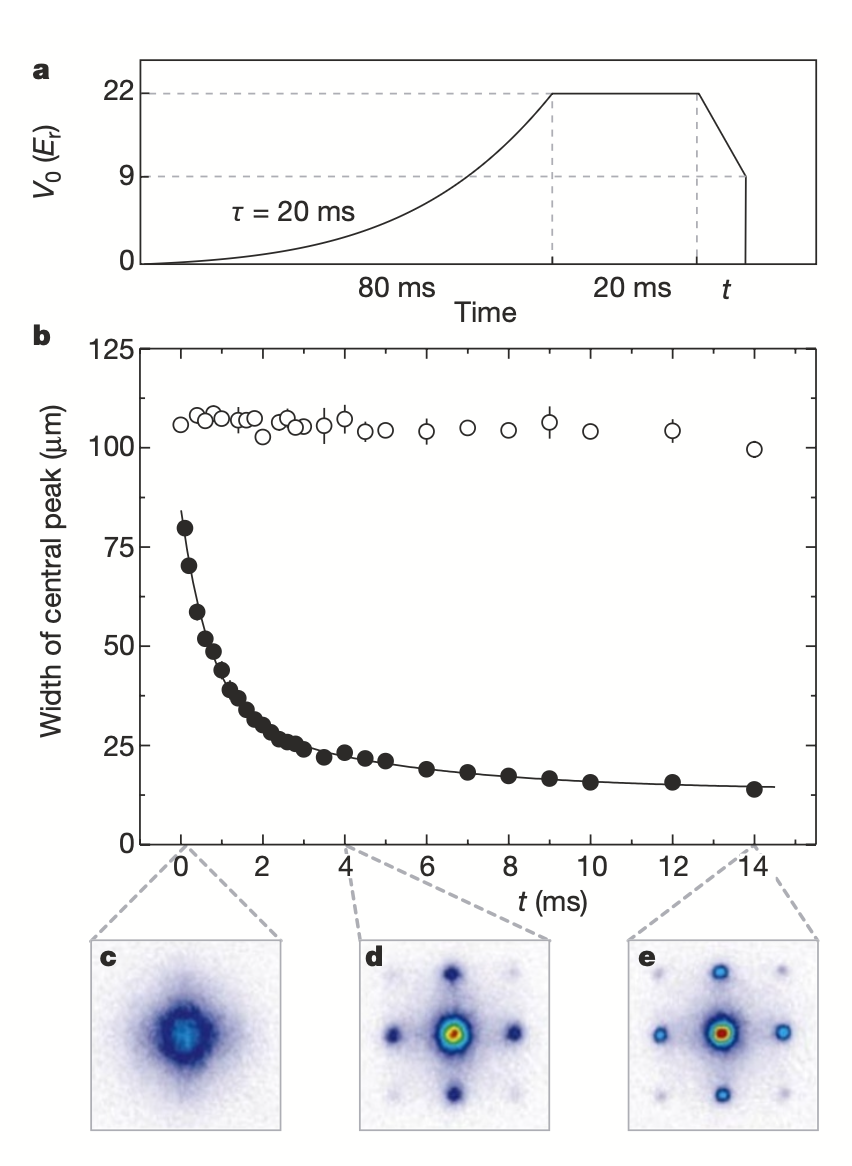} 
\caption{Restoring coherence. \textbf{a}, Experimental sequence used to measure the restoration of coherence after bringing the system into the Mott insulator phase at $V_0 = 22E_r$ and lowering the potential afterwards to $V_0 = 9E_r$, where the system is superfluid again. The atoms are first held at the maximum potential depth $V_0$ for 20 ms, and then the lattice potential is decreased to a potential depth of $9E_r$ in a time $t$ after which the interference pattern of the atoms is measured by suddenly releasing them from the trapping potential. \textbf{b}, Width of the central interference peak for different ramp-down times $t$, based on a lorentzian fit. In case of a Mott insulator state (filled circles) coherence is rapidly restored already after 4\,ms. The solid line is a fit using a double exponential decay ($\tau_1 = 0.94(7)$\,ms, $\tau_2 = 10(5)$\,ms). For a phase incoherent state (open circles) using the same experimental sequence, no interference pattern reappears again, even for ramp-down times $t$ of up to 400\,ms. We find that phase incoherent states are formed by applying a magnetic field gradient over a time of 10 ms during the ramp-up period, when the system is still superfluid. This leads to a dephasing of the condensate wavefunction due to the nonlinear interactions in the system. \textbf{c–e}, Absorption images of the interference patterns coming from a Mott insulator phase after ramp-down times $t$ of 0.1\,ms (\textbf{c}), 4\,ms (\textbf{d}), and 14\,ms (\textbf{e}).}
\label{fig:coherence_restoration}
\end{figure}

\subsection{Restoring coherence}

A notable property of the Mott insulator state is that phase coherence can be restored very rapidly when the optical potential is lowered again to a value where the ground state of the many-body system is completely superfluid (see Fig.~\ref{fig:coherence_restoration}). After only 4\,ms of ramp-down time, the interference pattern is fully visible again, and after 14\,ms of ramp-down time the interference peaks have narrowed to their steady-state value, proving that phase coherence has been restored over the entire lattice. The timescale for the restoration of coherence is comparable to the tunnelling time $\tau_{\text{tunnel}} = \hbar/J$ between two neighbouring lattice sites in the system, which is of the order of 2 ms for a lattice with a potential depth of 9\,$E_r$. A significant degree of phase coherence is thus already restored on the timescale of a tunnelling time.

It is interesting to compare the rapid restoration of coherence coming from a Mott insulator state to that of a phase incoherent state, where random phases are present between neighbouring lattice sites and for which the interference pattern also vanishes. This is shown in Fig.~\ref{fig:coherence_restoration}b, where such a phase incoherent state is created during the ramp-up time of the lattice potential (see Fig.\,\ref{fig:coherence_restoration} legend) and where an otherwise identical experimental sequence is used. Such phase incoherent states can be clearly identified by adiabatically mapping the population of the energy bands onto the Brillouin zones~\cite{greiner2001b,kastberg1995}. When we turn off the lattice potential adiabatically, we find that a statistical mixture of states has been created, which homogeneously populates the first Brillouin zone of the three-dimensional lattice. This homogeneous population proves that all atoms are in the vibrational ground state of the lattice, but the relative phase between lattice sites is random. Figure~\ref{fig:coherence_restoration}b shows that no phase coherence is restored at all for such a system over a period of 14 ms. Even for evolution times $t$ of up to 400 ms, no reappearance of an interference pattern could be detected. This demonstrates that the observed loss of coherence with increasing potential depth is not simply due to a dephasing of the condensate wavefunction.

\subsection{Probing the excitation spectrum}

\begin{figure}[htbp]
\centering
  \includegraphics[width=\linewidth]{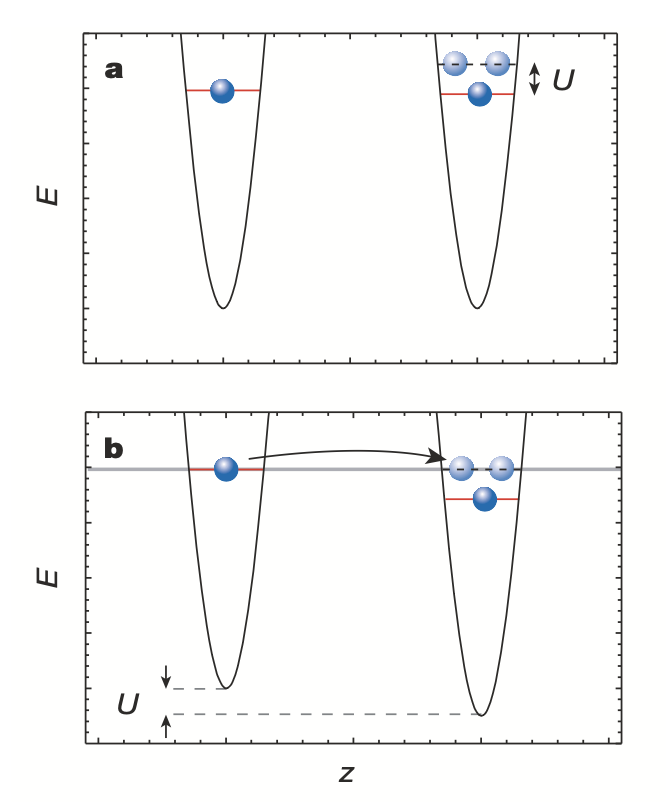} 
\caption{Excitation gap in the Mott insulator phase with exactly $n = 1$ atom per lattice site. \textbf{a}, The lowest lying excitations in the Mott insulator phase consist of removing an atom from a lattice site and adding it to neighbouring lattice sites. Owing to the on-site repulsion between the atoms, this requires a finite amount $U$ in energy and hopping of the atoms is therefore suppressed. \textbf{b}, If a potential gradient is applied to the system along the $z$-direction, such that the energy difference between neighbouring lattice sites equals the on-site interaction energy $U$, atoms are allowed to tunnel again. Particle–hole excitations are then created in the Mott insulator phase.}
\label{fig:excitation_gap}
\end{figure}

In the Mott insulator state, the excitation spectrum is substantially modified compared to that of the superfluid state. The excitation spectrum has now acquired an energy gap $\Delta$, which in the limit $J \ll U$ is equal to the on-site interaction matrix element $\Delta = U$ (see refs.\,\cite{sheshadri1993,freericks1995,oosten2001,elstner1999}). This can be understood within a simplified picture in the following way. We consider a Mott insulator state with exactly $n = 1$ atom per lattice site. The lowest lying excitation for such a state is the creation of a particle–hole pair, where an atom is removed from a lattice site and added to a neighbouring lattice site (see Fig.\,\ref{fig:excitation_gap}a). Due to the on-site repulsion between two atoms, the energy of the state describing two atoms in a single lattice site is raised by an amount $U$ in energy above the state with only a single atom in this lattice site.

Therefore in order to create an excitation the finite amount of energy $U$ is required. It can be shown that this is also true for number states with exactly $n$ atoms per lattice site. Here the energy required to make a particle-hole excitation is also $U$. Hopping of particles throughout the lattice is therefore suppressed in the Mott insulator phase, as this energy is only available in virtual processes. If now the lattice potential is tilted by application of a potential gradient, tunnelling is allowed again if the energy difference between neighbouring lattice sites due to the potential gradient equals the on-site interaction energy $U$ (see Fig. 4b). We thus expect a resonant excitation probability versus the applied energy difference between neighbouring lattice sites for a Mott insulator phase.

\begin{figure}[htbp]
\centering
  \includegraphics[width=\linewidth]{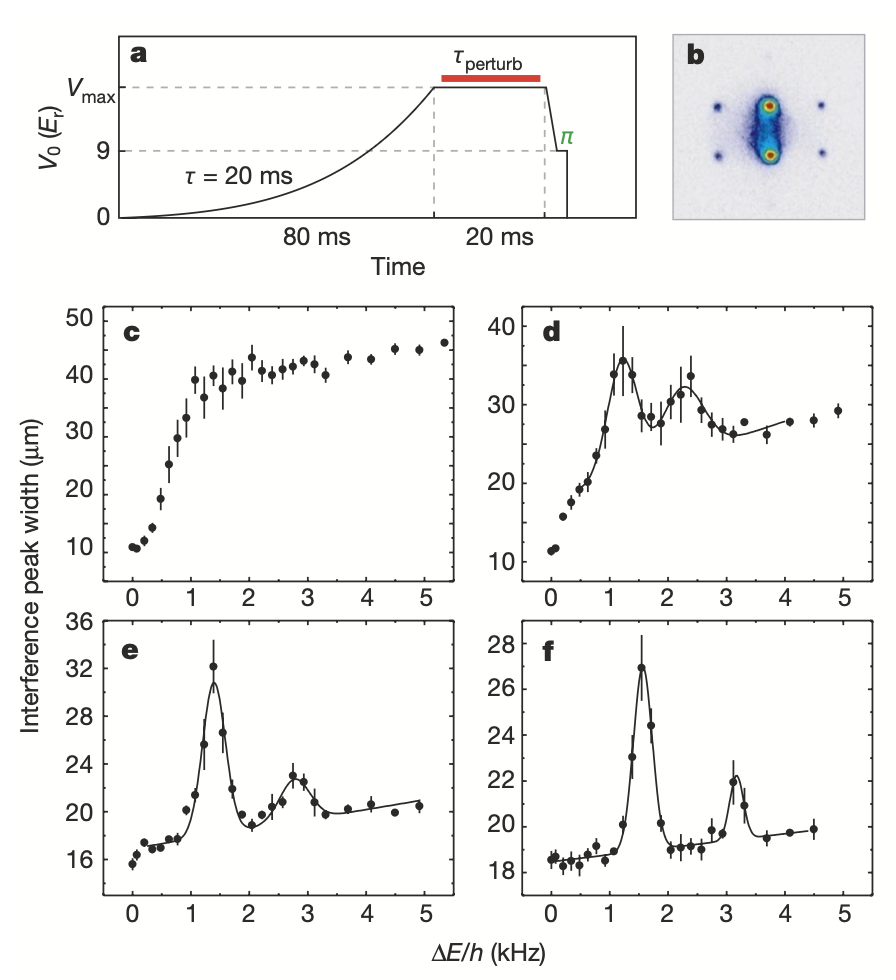} 
\caption{Probing the excitation probability versus an applied vertical potential gradient. \textbf{a}, Experimental sequence. The optical lattice potential is increased in 80\,ms to a potential depth $V_0 = V_{\max}$. Then the atoms are held for a time of 20\,ms at this potential depth, during which a potential gradient is applied for a time $\tau_{\text{perturb}}$. The optical potential is then lowered again within 3\,ms to a value of $V_0 = 9\,E_r$, for which the system is superfluid again. Finally, a potential gradient is applied for 300\,$\mu$s with a fixed strength, such that the phases between neighbouring lattice sites in the vertical direction differ by $\pi$. The confining potential is then rapidly turned off and the resulting interference pattern is imaged after a time of flight of 15 ms (\textbf{b}). Excitations created by the potential gradient at a lattice depth of $V_0 = V_{\max}$ will lead to excitations in the superfluid state at $V_0 = 9\,E_r$. Here excitations correspond to phase fluctuations across the lattice, which will influence the width of the observed interference peaks. \textbf{c–f}, Width of interference peaks versus the energy difference between neighbouring lattice sites $\Delta E$, due to the potential gradient applied for a time $\tau_{\text{perturb}}$. \textbf{c}, $V_{\max} = 10E_r$, $\tau_{\text{perturb}} = 2$ ms; \textbf{d}, $V_{\max} = 13E_r$, $\tau_{\text{perturb}} = 6$ ms; \textbf{e}, $V_{\max} = 16E_r$, $\tau_{\text{perturb}} = 10$ ms; and \textbf{f}, $V_{\max} = 20E_r$, $\tau_{\text{perturb}} = 20$ ms. The perturbation times $\tau_{\text{perturb}}$ have been prolonged for deeper lattice potentials in order to account for the increasing tunnelling times. The solid lines are fits to the data based on two gaussians on top of a linear background.}
\label{fig:excitation_spectroscopy}
\end{figure}

We probe this excitation probability by using the experimental sequence shown in Fig.~\ref{fig:excitation_spectroscopy}a. If excitations have been created during the application of the potential gradient at the potential depth $V_0 = V_{\max}$, we will not be able to return to a perfectly coherent superfluid state by subsequently lowering the potential to a depth of $V_0 = 9E_r$. Instead, excitations in the Mott insulator phase will lead to excitations in the lowest energy band in the superfluid case. These excitations are simply phase fluctuations between lattice sites, and cause a broadening of the interference maxima in the interference pattern (see Fig.\,\ref{fig:excitation_spectroscopy}b). Figure \ref{fig:excitation_spectroscopy}c-f shows the width of the interference peaks versus the applied gradient for four different potential depths $V_{max}$. 

For a completely superfluid system at 10 $E_r$, the system is easily perturbed already for small potential gradients and for stronger gradients a complete dephasing of the wavefunctions leads to a saturation in the width of the interference peaks. At a potential depth of about 13 $E_r$ two broad resonances start to appear in the excitation spectrum, and for a potential depth of 20 $E_r$ a dramatic change in the excitation spectrum has taken place. Two narrow resonances are now clearly visible on top of an otherwise completely flat excitation probability. The slightly higher offset of the excitation probability for a deep optical lattice (Fig.\,\ref{fig:excitation_spectroscopy}e,f) compared to the initial width of the interference peaks in Fig.\,\ref{fig:excitation_spectroscopy}c is due to the fact that after 3\,ms ramp down time from a deep optical lattice, the system is still in the dynamical process of restoring coherence coming from the Mott insulator phase. For longer hold times this offset approaches almost the same initial width as in Fig.\,\ref{fig:excitation_spectroscopy}c, showing that we are not able to excite the system at all except for the two resonance gradients. At these large potential depths, the narrow resonances show that the energy gap $\Delta$ of the system, which is measured here as the minimum energy difference between neighbouring lattice sites for which the system can be perturbed, is almost equal to the centre position of the
resonance.

We have in fact found the Mott insulator state to be extremely
robust to external perturbations, such as a modulation of the
trapping potential or a modulation of the gradient potential, as
long as the resonance gradients are avoided. The first resonance can be directly attributed to the creation of single particle-hole excitations in the Mott insulator state, and directly proves that we have indeed entered the Mott insulator regime. The second, weaker resonance occurs at exactly twice the energy difference of the first, stronger resonance. It can most probably be attributed to at least one of the following processes: (1) simultaneous tunnelling of two particles in a Mott insulator phase with $n > 1$ atoms, (2) second-
order processes, in which two particle-hole pairs are created
simultaneously, with only one in the direction of the applied
gradient, and (3) tunnelling processes occurring between lattice
sites with $n=1$ atom next to lattice sites with $n=2$ atoms. In
comparison, a two-dimensional lattice at a maximum potential
depth of $V_{max} = 20\,Er$, which we still expect to be in the superfluid regime, shows no resonances but a smooth excitation spectrum, similar to Fig.\,\ref{fig:excitation_spectroscopy}c.

\begin{figure}[htbp]
\centering
  \includegraphics[width=\linewidth]{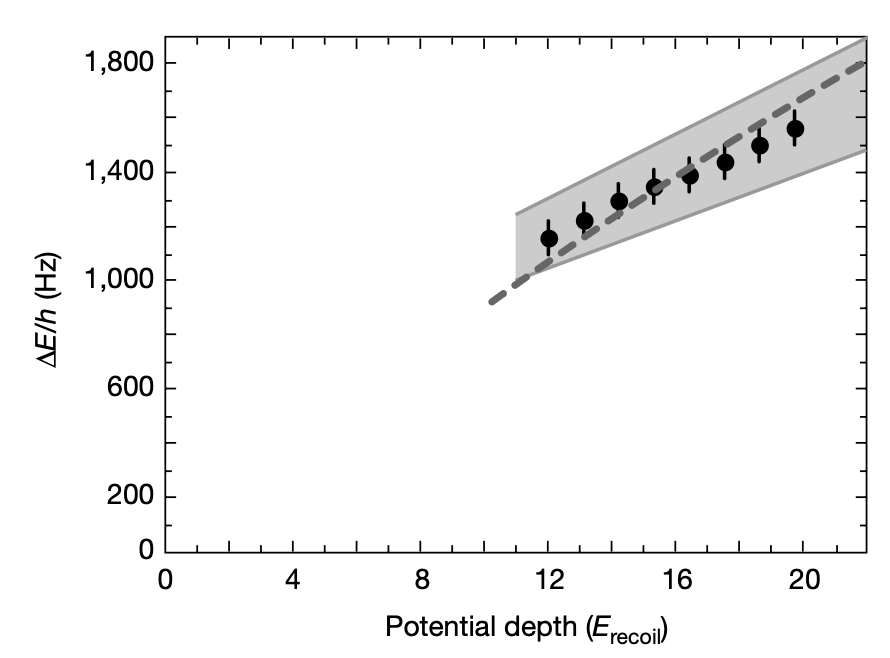} 
\caption{Energy difference between neighbouring lattice sites $\Delta E$ for which the Mott insulator phase can be resonantly perturbed versus the lattice potential depth $V_{\max}$. Experimental data points are shown as filled circles, and the shaded grey area denotes the possible variation of experimental values due to systematic uncertainties in the calibration of the potential depth and the applied gradient. The dashed line is the theoretical prediction for the on-site interaction matrix element $U$, based on Wannier functions from a band structure calculation.}
\label{fig:resonance_position}
\end{figure}

The position of the resonances in the three-dimensional lattice can be seen to shift with increasing potential depth due to the tighter localization of the wave packets on a lattice site (see Fig.~\ref{fig:resonance_position}). We have compared the position of the first resonance versus the potential depth $V_{\max}$ to an ab initio calculation of $U$ based on Wannier functions from a band structure calculation, and find good agreement within our experimental uncertainties (see Fig.\,\ref{fig:resonance_position}).

\section{Transition Point}

Both the vanishing of the interference pattern and the appearance of resonances in the excitation spectrum begin to occur at potential depths of $V_0 = 12(1)-13(1)\,E_r$, indicating the transition to the Mott insulator phase. We therefore expect the experimental transition point to lie above $V_0 = 10(1)\,E_r$, where no resonances are visible, and below $V_0 = 13(1)\,E_r$. It is important to compare this parameter range to the theoretical prediction based on the expected critical value $U/J = z \times 5.8$. In our simple cubic lattice structure, six next neighbours surround a lattice site. $J$ and $U$ can be calculated numerically from a band structure calculation for our experimental parameters, from which we find that $U/J \approx 36$ for a potential depth of 13 $E_r$. The theoretical prediction for the transition point is therefore in good agreement with the experimental parameter range for the transition point.

\section{Outlook}

We have realized experimentally the quantum phase transition from a superfluid to a Mott insulator phase in an atomic gas trapped in an optical lattice. The experiment enters a new regime in the many-body physics of an atomic gas. This regime is dominated by atom–atom interactions and it is not accessible to theoretical treatments of weakly interacting gases, which have so far proved to be very successful in describing the physics of Bose–Einstein condensates~\cite{dalfovo1999}.

The experimental realization of the Bose–Hubbard model with an atomic gas now allows the study of strongly correlated many-body quantum mechanics with unprecedented control of parameters. For example, besides controlling mainly the tunnelling matrix element, as done in this work, it should be possible in future experiments to control the atom–atom interactions via Feshbach resonances~\cite{inouye1998,donley2001}.

The atoms in the Mott insulator phase can be considered as a new state of matter in atomic gases with unique properties. Atom number fluctuations at each lattice site are suppressed, and a well-defined phase between different lattices sites no longer exists. These number states have been proposed for the realization of a Heisenberg-limited atom interferometer~\cite{bouyer1997}, which should be capable of achieving improved levels of precision. The Mott insulator phase also opens a new experimental avenue for recently proposed quantum gates with neutral atoms~\cite{jaksch1999}.

\acknowledgments
We thank W. Zwerger, H. Monien, I. Cirac, K. Burnett and Yu. Kagan for discussions. This work was supported by the DFG, and by the EU under the QUEST programme.

\end{document}